\documentclass[twocolumn,twoside,slac_two]{revtex4}
\usepackage{graphicx}
\usepackage{fancyhdr}
\begin{document}

\title{The Optical Properties of PKS 1222+216 During the Fermi Mission}

%

\author{P. S. Smith}
\affiliation{Steward Observatory, University of Arizona, Tucson, AZ 85716 USA}
\author{G. D. Schmidt}
\affiliation{National Science Foundation, Arlington, VA 22230 USA}
\author{B. T. Jannuzi}
\affiliation{NOAO, Kitt Peak National Observatory, Tucson, AZ 85726 USA}

\begin{abstract}
The optical properties of the $z = 0.435$ quasar PKS 1222+216
(4C+21.35) are summarized since the discovery
of impressive $\gamma\/$-ray activity in this source by {\it Fermi\/}/LAT. Unlike several
other $\gamma\/$-ray-bright blazars, there appears to be little connection
between optical and $\gamma\/$-ray activity. Spectropolarimetry
shows this object to be a composite system with optical
emission from both a polarized, variable synchrotron power-law
and unpolarized light from a stable blue continuum source (+broad
emission-line region) contributing to the observed spectrum.
Spectrophotometry over a period of about two years does
not detect significant variability in the strong, broad emission lines,
despite large optical continuum variations.
This suggests that the relativistic jet has little influence on the 
output of the broad emission-line region, possibly either because the highly beamed 
continuum ionizes only a small portion of the line-emitting gas, or the
observed non-thermal continuum originates parsecs downstream 
from the base of the jet, further away from the central engine than 
the broad emission-line region. 

\end{abstract}

\maketitle

\thispagestyle{fancy}


\section{INTRODUCTION AND OBSERVATIONS}
Since the announcement on 2009 April
17~\cite{ref1} that the Large Area Telescope (LAT) aboard the 
{\it Fermi Gamma-ray Space Telescope\/}~\cite{lat} detected increased $\gamma\/$-ray emission from PKS 1222+216, we have been systematically monitoring this $z = 0.435$ blazar at
optical wavelengths.
PKS 1222+216 is now among the two dozen or so $\gamma\/$-ray-bright blazars that
form the core of the sample being monitored by Steward
Observatory~\cite{ref2}
in support of {\it Fermi\/}. This optical program uses the 2.3~m Bok
and 1.54~m Kuiper telescopes with the SPOL spectropolarimeter~\cite{ref3} and provides
publicly available spectropolarimetry, spectrophotometry, and calibrated broad-band flux measurements
for about 40 blazars.\footnote{http://james.as.arizona.edu/$\sim$psmith/Fermi/} 
PKS 1222+216
provides an example of how these data can be used to construct a comprehensive view
of the optical behavior of blazars that can then be compared to observations made by the LAT and at
other wavelengths.

In both the $\gamma\/$-ray and optical spectral regimes, the variability of PKS 1222+216 is
characterized by numerous short-duration flares. For
instance, the 0.1-300 GeV light curve shows at least 8
outbursts, each lasting just a few days at most (Figure~1). Similarly,
large daily fluctuations in both the optical flux and
polarization are observed over the $\sim$2-year period. Interestingly, the optical and $\gamma\/$-ray
activity do not show a direct correspondence, unlike several
other well-studied blazars where the site(s) of $\gamma\/$-ray
production can be directly tied to those producing the bulk of
the optical and radio flux~\cite{ref4}, \cite{ref5}, \cite{ref6}. In particular, a recent
optical outburst of PKS 1222+216 in 2011 March occurred
without any apparent corresponding $\gamma\/$-ray activity. 

\begin{figure}
\includegraphics[width=85mm]{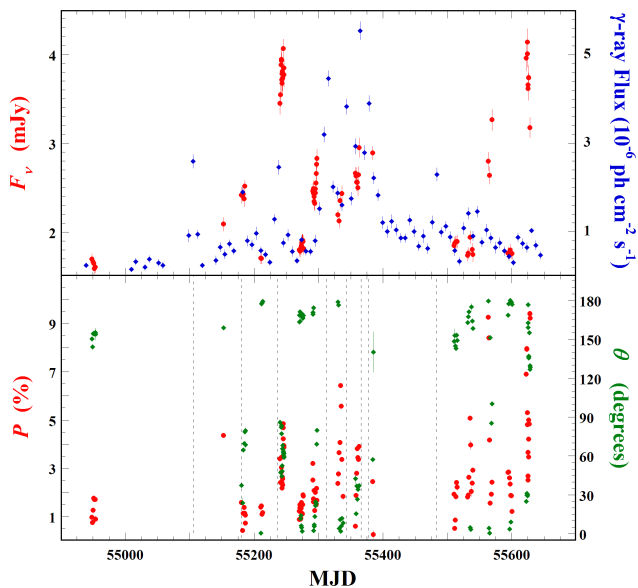}
\caption{{\it Top panel\/}: The $\gamma\/$-ray (blue)
and optical (red; from 2009 April 27--
2011 March 8) flux variations of PKS
1222+216. Weekly LAT averages are
shown for the $\gamma\/$-ray light curve.
{\it Bottom panel\/}: Optical polarization
variations, with the degree of
polarization, $P\/$, shown in red and the
polarization position angle, $\theta\/$, in
green. The dotted vertical lines denote
the occurrence of eight distinct $\gamma\/$-ray
outbursts. The major optical outburst
during 2011 March (MJD $\sim$ 55625) has
no apparent $\gamma\/$-ray analog.}
\label{l2ea4-f1}
\end{figure}

\begin{figure*}
\includegraphics[width=100mm]{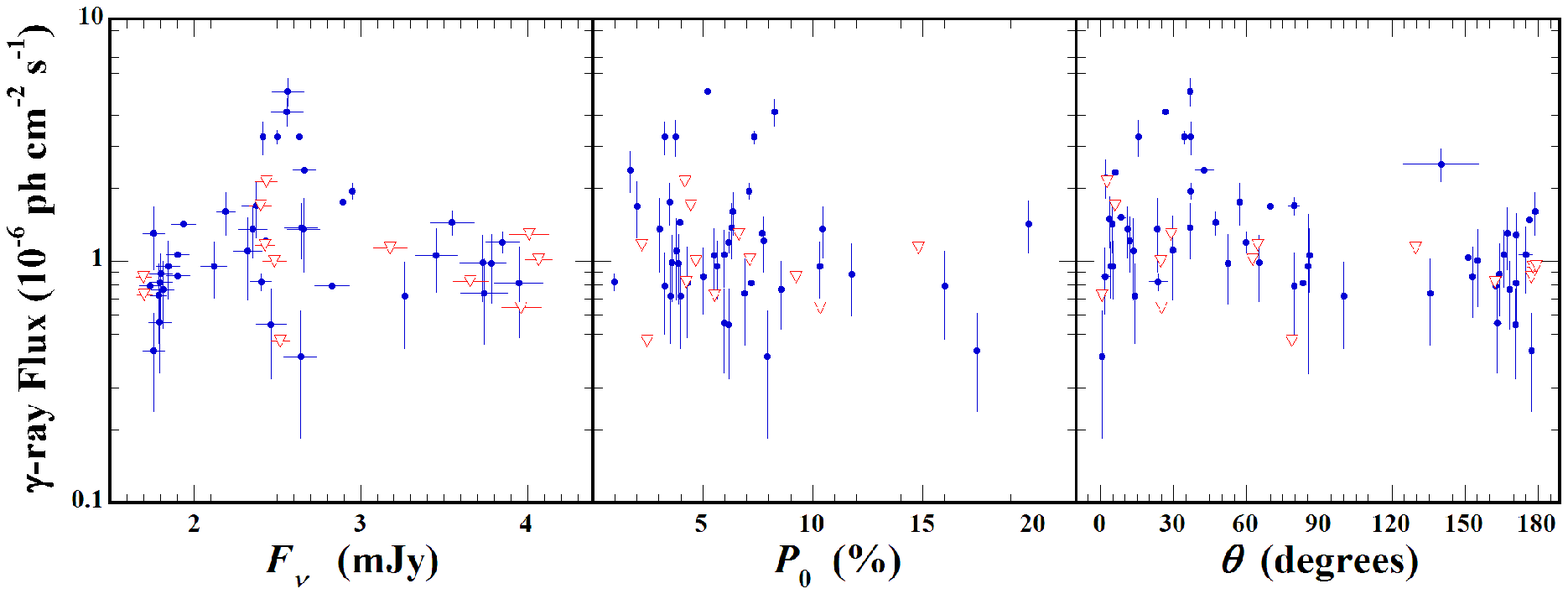}
\caption{Simultaneous $\gamma\/$-ray and
optical measurements. No clear trend
is observed between $\gamma\/$-ray flux and
either optical flux, polarization (corrected for unpolarized emission; see \S3), or the polarization position angle, $\theta\/$.
Upper limits for single-day LAT fluxes
are shown as red triangles.}
\label{l2ea4-f2}
\end{figure*}

Figure~2 summarizes the nearly simultaneous (within 12~hr) LAT $\gamma\/$-ray and
optical measurements from Steward Observatory of PKS 1222+216.
The $\gamma\/$-ray flux is not found to be correlated with the $V\/$-band optical brightness, the degree of 
optical polarization of the synchrotron continuum ($P_0\/$; see \S3),
or the polarization position angle, $\theta\/$.
Although nearly the entire range of $\theta\/$ is observed during monitoring 
program, the majority of the optical
measurements show $\theta\/$ 
to be within 30$^\circ\/$ of north, more-or-less
aligned with both the position angles of the VLBI jet~\cite{ref7}, \cite{ref8} and the
polarized flux of the millimeter core~\cite{ref8}.

\section{EMISSION LINES}

The optical spectrum of the object exhibits strong, broad
MgII and Balmer emission lines. Narrow-line emission in PKS 1222+216 is weak,
although [O~III]$\lambda\/$5007 is detected in the moderate-resolution
spectra obtained by SPOL.
Because the Steward Observatory program provides calibrated 
spectrophotometry in addition to linear polarization data, the flux
spectrum of PKS 1222+216 is monitored routinely.
The emission from the broad-line region does
not vary on the extremely short time scales observed for the
continuum.
Indeed, measurements of the H$\beta\/$ and H$\gamma\/$ fluxes over the two-year
period do not detect significant changes in the emission-line flux (Figure~3, {\it right panel\/}).  The constancy of the line fluxes is also reflected in the {\it left
panel\/} of Figure~3, as the equivalent widths (EWs) of both H$\beta\/$ and H$\gamma\/$
systematically decrease as the optical continuum brightens.
The larger scatter in the measurements for H$\beta\/$ compared to those for
H$\gamma\/$ are due to the presence of the strong atmospheric O$_2$ absorption 
feature in the blue wing of H$\beta\/$ and fringing of the thinned CCD at these 
wavelengths.

\begin{figure*}
\includegraphics[width=110mm]{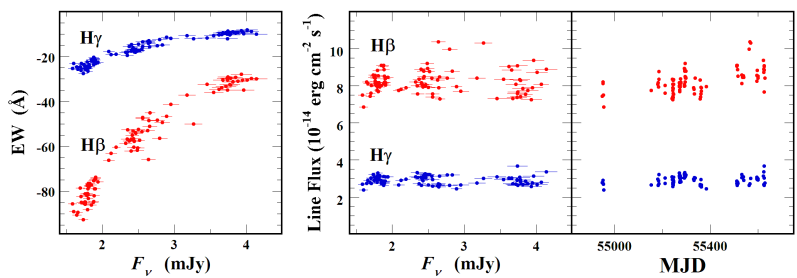}
\caption{Measurements of the
equivalent width (EW) and flux of the
H$\beta\/$ and H$\gamma\/$ emission lines. The line
fluxes stay relatively constant with
time and continuum brightness over
the $\sim$2-yr period. As a result, the EWs
are closely tied to variations of the
optical continuum.}
\label{l2ea4-f3}
\end{figure*}

The stability seen in the emission lines is consistent with the broad-line region
of the blazar not being affected by the highly beamed and variable continuum 
produced by the jet.  The lack of a coupling between the jet and the broad emission-line
region may be the result of the beamed continuum intersecting only a small fraction
of the volume 
containing the emission-line gas.  The observed jet emission could also originate further
away from the central engine than the extent of the region containing the broad-line emitting clouds, which
is measured to be a few to several light months from the ionizing continuum source
(see, e.g.,~\cite{ref10}). 
The latter possibility is in line with other evidence that suggests that the 
non-thermal continuum originates several parsecs from the base
of the relativistic jet~\cite{ref5}, \cite{ref6}, \cite{ref11}.

\section{OPTICAL CONTINUUM AND POLARIZATION}
Although few connections can be found between the optical and $\gamma\/$-ray variability in PKS 1222+216, strong correlations are found optically between flux, color, and
polarization (Figure~4). As PKS 1222+216 brightens, the continuum becomes redder. In addition, the polarization
is almost always observed to decrease toward the blue regardless of the level of
polarization observed ($P_{obs} < 10$\%). The decrease in $P\/$ to the blue
is also generally found to be stronger when the blazar is faint ({\it middle panel\/} of Figure~4).
Similarly, the polarization is observed to decrease in the major broad emission lines.
This is especially apparent for the high-EW H$\beta\/$ line, and is consistent
with the emission-line flux being completely unpolarized.
In contrast to $P\/$, the polarization position angle generally remains constant 
across the spectrum. 
These trends are well illustrated in the individual observations of PKS 1222+216
shown in Figure~5.

\begin{figure*}
\includegraphics[width=110mm]{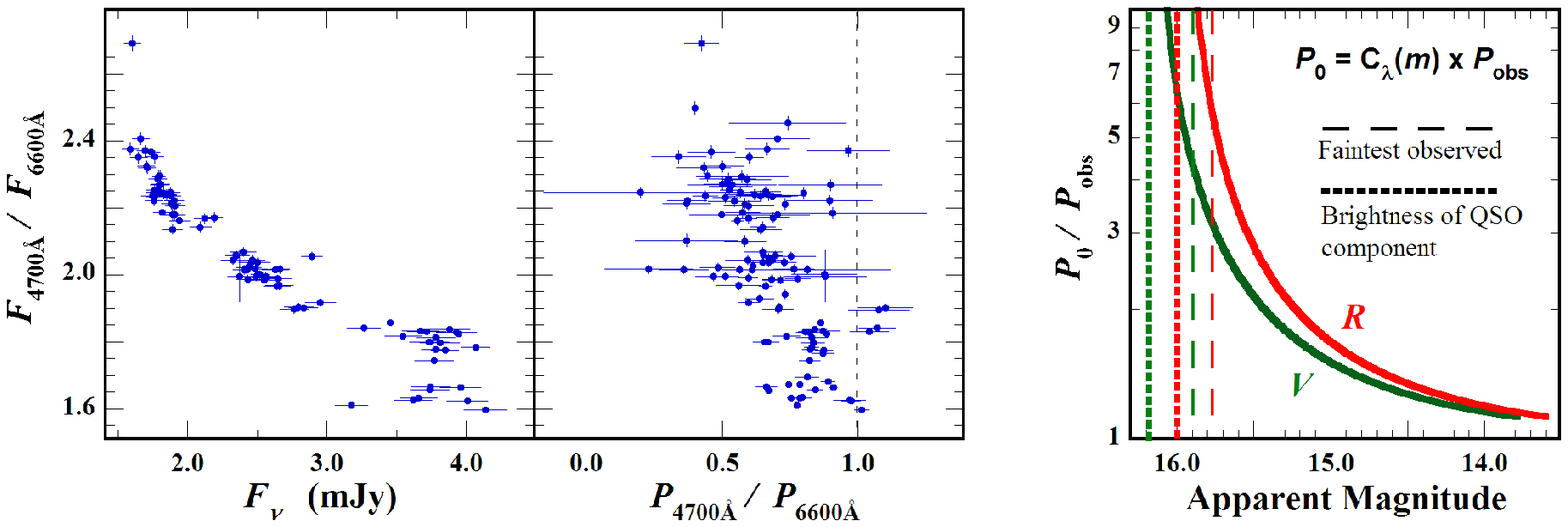}
\caption{{\it Left panel\/}: The flux density ratio between
4700~\AA\ and 6600~\AA\ plotted against the $V\/$-band
brightness. The object becomes redder as it
brightens. {\it Center panel\/}: The same quantity plotted
against the ratio of polarization at the same
wavelengths. As the object fades, the polarization
decreases more strongly to the blue. {\it Right panel\/}:
The correction to the observed broad-band
polarization as a function of apparent magnitude,
assuming a constant source of light that dilutes the
polarized flux ($P_0 \times F_{\nu}\/$) from a power-law continuum.}
\label{l2ea4-f4a}
\end{figure*}

\begin{figure}
\includegraphics[width=65mm]{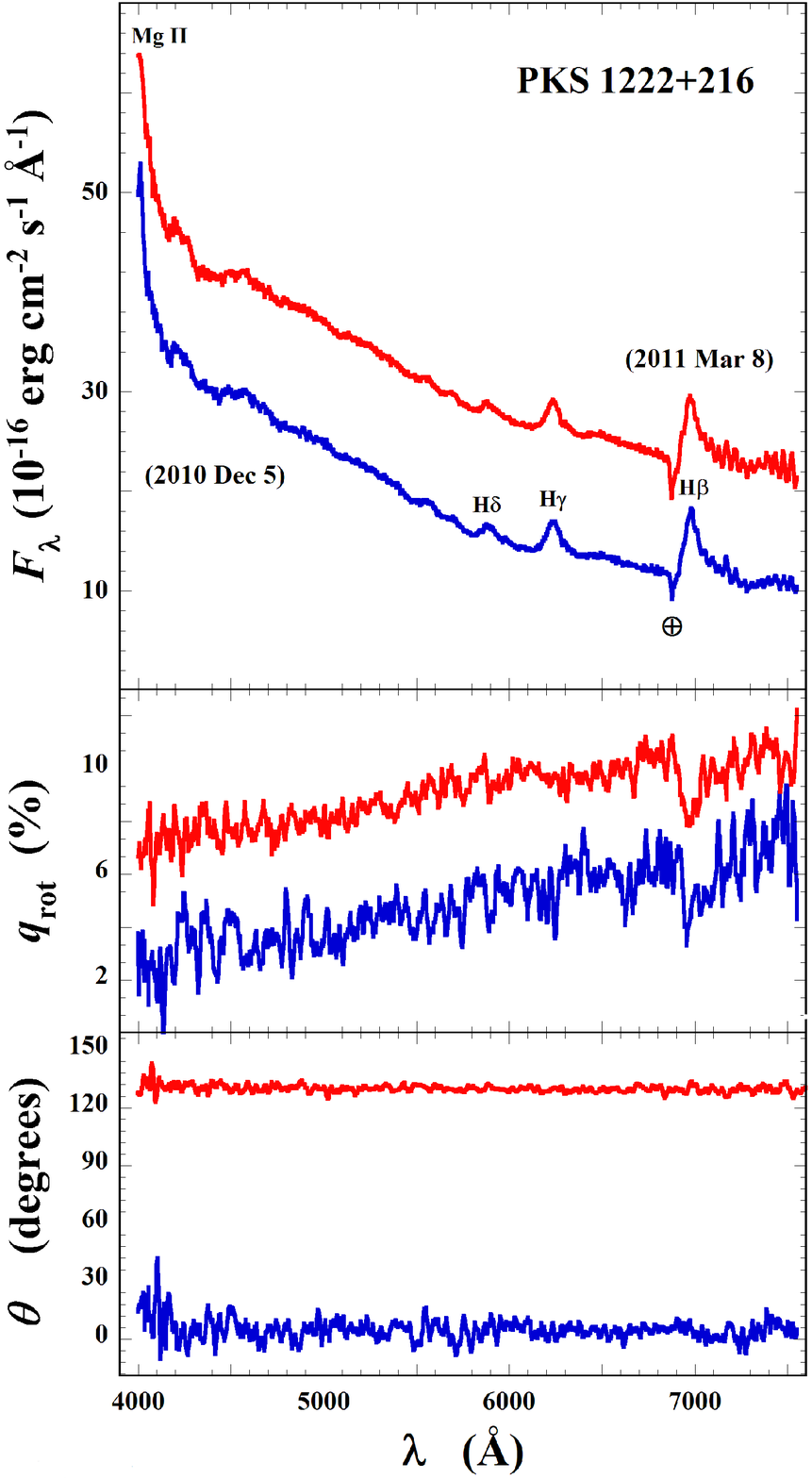}
\caption{Spectropolarimetry in the observed reference frame when the blazar is bright (red) and
faint (blue). {\it Top panel\/}: The spectrum with major emission lines
identified. The atmospheric O$_2$ feature in the blue wing of H$\beta\/$ is also
marked. {\it Middle panel\/}: The $q\/$ Stokes parameter rotated so that $u\/$
averages to 0 over the spectrum. There is a distinct decrease in polarization
to the blue when the object is either bright or faint. {\it Bottom panel\/}: The spectrum of $\theta\/$, which is generally
observed to be constant with wavelength, as expected if the polarized
emission is dominated by a single non-thermal source.}
\label{l2ea4-f4b}
\end{figure}

The polarization properties and the correlation between brightness and optical
color lead directly to a picture that has the optical emission coming from two
sources: (1) a variable, polarized synchrotron continuum that is produced by the
relativistic jet, and (2) a much more stable (at least on the time scale of a year
or more), unpolarized source with a spectrum similar to an optically-selected
QSO.  Figure~6 shows an illustrative example of how such a simple two-component
system explains all of the major optical correlations seen in PKS 1222+216 during
the monitoring program.
In this model, the synchrotron continuum is well described by a simple power law
having constant polarization over the observed spectral range.
Fitting the wavelength dependence observed in $P\/$ is accomplished by assuming
that the flux additional to the synchrotron continuum is unpolarized.
The resulting flux spectrum that explains the variation in polarization as a 
function of wavelength is quite similar to a typical optically-selected QSO
in continuum color and broad-line emission.  Similar models have been very successful in 
explaining the optical properties of several other highly polarized quasars 
(see~\cite{smith}, and references therein).

The particular model for PKS 1222+216 on 2011 March~8 (Figure~6) reveals that the intrinsic
polarization ($P_0\/$) of the power-law component is independent of wavelength at nearly 
15\% and that its 
spectral index is $\alpha \sim -1.3$.  Furthermore, the correlations between 
brightness, color, and polarization are reproduced by the model if the 
QSO component (essentially, a ``Big Blue Bump'' continuum that likely provides the
ultraviolet flux that ionizes the broad emission-line region)
is kept constant.
The flux-color correlation (see Figure~4) arises because as the variable jet emission
brightens, its relatively redder continuum becomes more dominant and reddens the overall
spectrum.
The models and observed correlations suggest an unpolarized component with an 
apparent $V\/$ magnitude $\sim$16.2 ({\it right panel\/} of Figure~4).
In turn, the observed polarizations can be corrected by subtracting the unpolarized light from the total flux, yielding an estimate of the intrinsic polarization of the
synchrotron continuum.  The corrected polarizations are shown in the 
{\it middle panel\/} of Figure~2 and can be substantial.  The maximum observed 
polarization during period monitored is shy of 10\%, but $P_0\/$ can reach
$\sim$20\%, putting PKS 1222+216 on par with other blazars.  Of course, the
dilution of the non-thermal polarization by the QSO component has no effect on $\theta\/$.

\begin{figure}
\includegraphics[width=65mm]{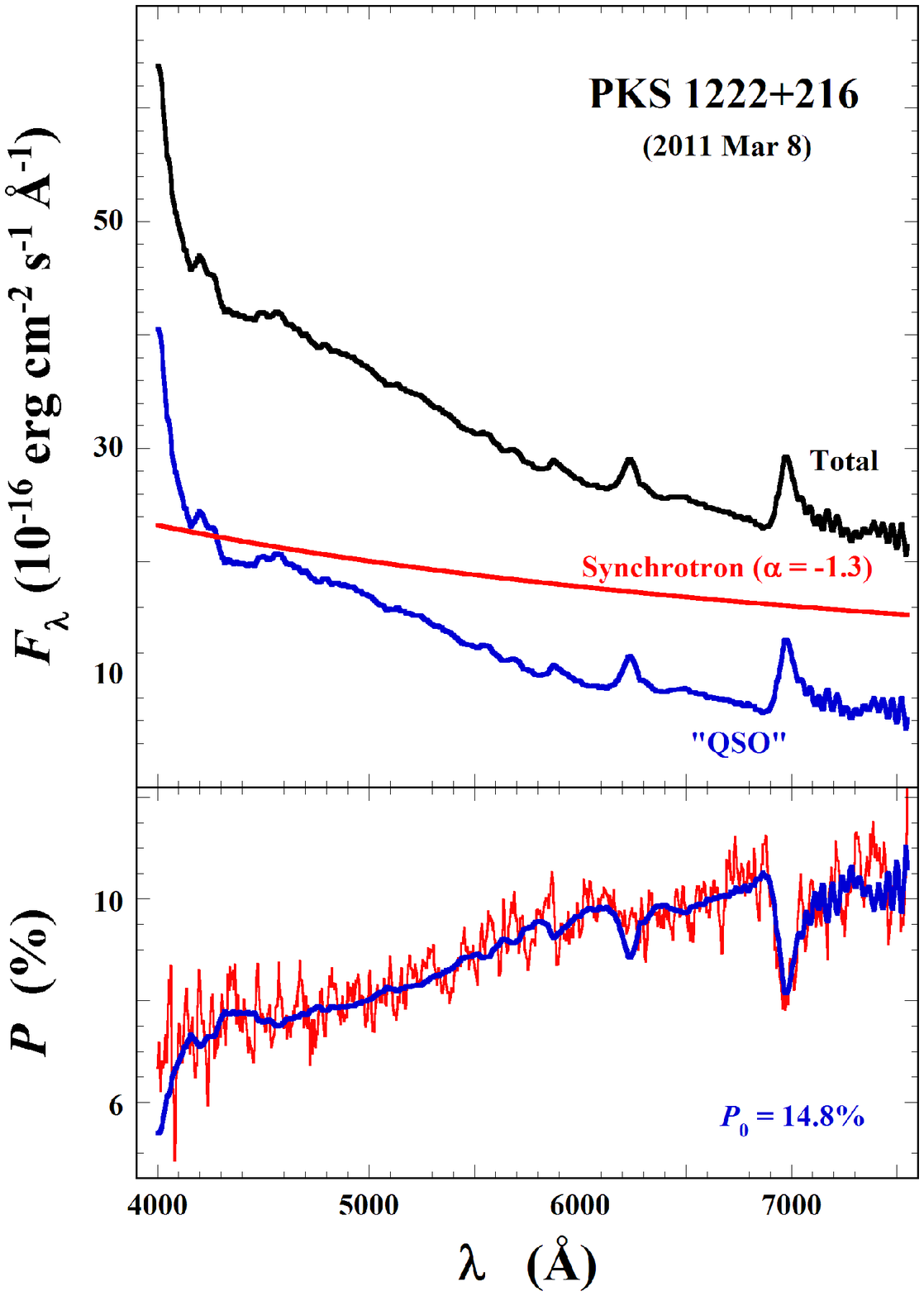}
\caption{PKS 1222+216 on 2011
March 8. {\it Top panel\/}: The optical
spectrum is modeled with a
polarized ($P_0\/$ = 14.8\%), power-law
synchrotron source and an
unpolarized, continuum+broad
emission-line spectrum similar to a
typical optically-selected QSO.
{\it Bottom panel\/}: The effect of the
dilution of the non-thermal
polarized flux by the bluer QSO
component. The model is shown
by the blue curve and the
observed data indicated in red.
Such a picture explains both the
decrease in $P\/$ toward shorter
wavelengths and the observed
correlation between optical color and flux.}
\label{l2ea4-f5}
\end{figure}

\section{SUMMARY AND CONCLUSIONS}

PKS 1222+216 is an important object for continued intensive study across the
entire electromagnetic spectrum. Although direct connections can be made
between the optical and radio emission owing to the rough alignment of the
polarizations and the position angle of the inner VLBI jet, there are
significant differences between the behavior of this quasar relative to other $\gamma\/$-ray-bright blazars at optical and GeV energies. In particular, major optical or $\gamma\/$-ray
events can occur with no obvious corresponding activity in the other energy
regime. The importance of PKS 1222+216 to the questions of the how and where
high-energy photons are produced in blazars is underlined by the recent
discovery of TeV emission from this source~\cite{ref9}. Continued optical spectropolarimetry enables not only efficient monitoring of the high-energy tail of the
primary relativistic electrons and the magnetic field within the jet, but also the
``normal QSO'' emission, which provides valuable information on the accretion
processes that presumably give rise to the jet.

\begin{acknowledgments}
Monitoring of PKS 1222+216 and $\sim$40 other $\gamma\/$-ray-bright blazars at
Steward Observatory has
been made possible by NASA {\it Fermi\/} Guest Investigator grants 
NNX08AV65G and NNX09AU10G.
\end{acknowledgments}



\end{document}